Title: A 454 survey of the community composition and core microbiome of the common bed bug, *Cimex lectularius*, reveals significant microbial community structure across an urban landscape

Matthew Meriweather,[1] Sara Matthews,[1] Rita Rio[2], and Regina S Baucom[1,3]

[1]721 Rieveschl Hall, Department of Biological Sciences, University of Cincinnati, Cincinnati, OH 45221
[2]53 Campus Drive, Department of Biology, West Virginia University, Morgantown, WV 26506
[3]Author for correspondence: regina.baucom@uc.edu

**Abstract**

Elucidating the spatial dynamic and core constituents of the microbial communities found in association with arthropod hosts is of crucial importance for insects that may vector human or agricultural pathogens. The hematophagous *Cimex lectularius*, known as the common bed bug, has made a recent resurgence in North America, as well as worldwide, potentially owing to increased travel and resistance to insecticides. A comprehensive survey of the bed bug microbiome has not been performed to date, nor has an assessment of the spatial dynamics of its microbiome. Here we present a survey of bed bug microbial communities by amplifying the V4-V6 hypervariable region of the 16S rDNA gene region followed by 454 Titanium sequencing using 31 individuals from eight natural populations collected from residences in Cincinnati, OH. Across all samples, 97% of the microbial community is made up of two dominant OTUs identified as the α-proteobacterium *Wolbachia* and an unnamed γ-proteobacterium from the Enterobacteriaceae. Microbial communities varied among host populations for measures of community diversity and exhibited significant population structure. We also uncovered a strong negative correlation in the abundance of the two dominant OTUs, suggesting they may fulfill similar roles as nutritional mutualists. This broad survey represents the most comprehensive assessment, to date, of the microbes that associate with bed bugs, and uncovers evidence for potential antagonism between the two dominant members of the bed bug microbiome.

**Introduction**

The interactions between microbes and their insect hosts can range from beneficial to parasitic and ultimately detrimental to host fitness [1,2,3]. The bacteria that insects carry can also negatively impact other organisms following vector transmission. Characterizing the microbial community of insect hosts is thus an important primary step for elucidating the nature of host/microbe interactions and the potential for the insect vectoring of important pathogens. Assessments of the core microbiome—defined as members of the microbial community that are common to two or more assemblages of a particular habitat type [4,5]—are often performed to identify key host/microbe associations. The microbes that appear in all hosts are thought to fulfill a functional niche within the community [6] and as such may provide valuable information on the 'normal' state of the community as well as on what members might be targeted for manipulation (e.g., in a scenario of insect biocontrol).

The microbiome of an insect host could vary among host populations, and such natural variation would represent one aspect of an ecological deviation from the core microbiome [6]. Although assessments of the microbiome of various arthropods is an area of very active research [7,8,9,10,11,12,13,14,15], the potential for natural, spatial variation in the microbiome has been elucidated in only a handful of insect species [8,10,16]— perhaps most thoroughly in the model species *Drosophila* [9,17]. Interrogating spatial variation in the core microbiome of a range of insects, and specifically arthropods that may vector disease is of interest for three main reasons: first, the maintenance of a cohort of microbial species across many insect populations may provide information about important components of host physiology; second, the potential for interaction between endosymbiont species may be elucidated by wide-ranging surveys of microbial taxon abundance and frequency of co-occurrence, and finally, surveys of the spatial variation in microbiomes may identify populations in regions of importance to human and/or agricultural health that may then be targeted for remediation.

*Cimex lectularius*, the bed bug, has made a now infamous resurgence in North America as well as worldwide [18,19,20,21]. Bed bug infestations, while not restricted to social or economic groups, can disproportionately impact the economically disadvantaged since the efficacy of less expensive methods of control, such as pesticide application, is low owing to widespread pyrethroid resistance [22,23]. While there is currently no evidence that bed bugs vector disease, human pathogens have been identified on the surface of the insects and in their excrement [24]. Further, the presence of this hematophagous insect in dwellings can lead to skin reactions, anaphylaxis, anxiety and psychological problems in humans [24,25,26], and thus bed bugs are of broad, and increasing concern from the standpoint of human health [27].

Endosymbiont microbial species in *C. lectularius* were first identified by microscopy in 1921 [28], and recent 16S rRNA surveys have identified two main endosymbionts of the bed bug to be *Wolbachia* [29,30,31] and an unnamed, γ-proteobacterium closely related to an endosymbiont species from the leafhopper *Euscelidius variegatus* (termed BEV-like symbiont [29]. A high prevalence of *Wolbachia* infections has been found in North American populations of *C. lectularius* (between 83-100%, [32], and 16S rRNA sequencing has identified the strain to belong to the F supergroup [30,32,33]. Unlike the B supergroup of *Wolbachia* that is found to parasitize many arthropods, the F supergroup is not known to harbor any reproductive parasites [33]. Recent work has shown that *Wolbachia* is a nutritional mutualist of *C. lectularius* and provides the species with B vitamins essential for growth and reproduction. For this reason, *Wolbachia* has been proposed as an obvious target for the biocontrol of this insect pest [33]. However, before such an endeavor can be expected to succeed, more information regarding the natural microbial community of *C. lectularius* is needed. A broad assessment and characterization of the core microbiome of bed bugs collected from an urban area is thus highly warranted.

Here, we surveyed the microbiome of natural populations of bed bugs collected from residences located in metropolitan Cincinnati, OH (Fig 1), which is currently listed as the second most bed bug infested US city [34,35]. To make our assessment comprehensive, we surveyed between 2 and 5 *C. lectularius* individuals from each of eight geographically separated populations. We used culture-independent 16S rDNA PCR amplification followed by 454 Titanium sequencing to characterize the microbial community associated with each *C. lectularius* individual. We performed this survey to broadly characterize the core microbiome of naturally collected bed bug populations and to assess the following: Do the microbial communities of *C. lectularius* exhibit significant population structure? Are there differences among the populations in microbial community diversity and/or membership? What are the patterns of abundance across populations of the dominant endosymbionts *Wolbachia* and the unnamed γ-proteobacterium? Is there any evidence for pathogenic microbial groups within populations of *C. lectularius* collected from local residences?

**Results**

**Data summary**

Cimex samples were collected from residences in the greater Cincinnati metropolitan area in Cincinnati, OH, USA, by pest control technicians and transported to the Baucom lab (Fig 1). Between 10 and 50 individuals were collected per population in a sterile 50 mL falcon tube and starved for a week prior to being snap frozen in liquid $N_2$. Five individuals from each of the eight populations were randomly sampled and weighed in sterile conditions prior to DNA extraction. Individuals were not surface sterilized prior to isolation since we were interested in understanding the broad population dynamics of the microbial community found in association with wild *Cimex* populations. Following DNA extraction, the hypervariable V4-V6 region of the bacteria-specific 16S rDNA gene was PCR amplified in triplicate for each individual. Libraries from the same individual were then combined, cleaned, normalized, and sequenced using 454 Titanium chemistry.

From the initial 69,854 sequences, 11,147 were removed for being <200bp or low quality, 3,927 were identified as chimeric, and 666 were identified as either mitochondrial or chloroplast. After filtering the sequence reads for quality scores, sequencing errors and chimeras, our dataset consisted of 54,114 sequences distributed across 31 *C. lectularius* individuals (also referred to as libraries)(STable1) with an average read length of 284 bases. The 31 libraries varied in size from 74 to 4092 sequences, and averaged 1745.6 ± 171.2 sequences. Most libraries (25) contained at least 1000 sequences. Clustering the data with mothur [36] at the 3% level of sequence divergence returned 359 OTUs across all libraries. Two OTUs overwhelmingly dominated the sequence data, with one OTU containing 34,354 sequences and the other prominent OTU containing 18,085 sequences. On average, OTUs contained 150 sequences, but many OTUs (239) were represented by 1 or 2 sequences. Twenty-nine OTUs were represented by 10 or more sequences.

**Bacterial taxa associated with wild *Cimex* populations**

Approximately 97% of all sequences within our dataset were assigned to two families within the Proteobacteria by the RDP classifier: Anaplasmataceae (63.48%) and the Enterobacteriaceae (33.57%). Two genera—*Wolbachia* and an unclassified genus from the Enterobacteriaceae (hereafter 'unnamed γ-proteobacterium')—comprised the majority of these families, at 63.48% and 33.42%, respectively (Table 1). The remaining OTUs cover roughly 3.10% of the data (STable3). All 31 individual *C. lectularius* libraries contained the dominant *Wolbachia* and the unnamed γ-proteobacterium; all eight populations had a shared OTU from each of the following genera at very low frequency: *Shinella*, *Sphingomonas*, *Methylobacterium*, *Pseudomonas*, a second unclassified genus from the Enterobacteriaceae, *Acinetobacter*, and *Cloacibacterium*, as well as a shared OTU from the suborder Propionibacterineae (Table 1 and STable2). Thus, a core microbiome in wild-collected populations of *Cimex* could be expected to contain *Wolbachia* and the γ-proteobacterium and possibly OTUs from the low-frequency groups listed above. We screened our data for putative pathogens that were previously found to associate with *Cimex* individuals, and while the length of the 16S fragment that we analyzed precludes the identification of bacteria at the level of the species, we detected five bacterial OTUs from the same genus as a potentially human-pathogenic bacterial species, while the remaining were invertebrate tropic (Table 2).

**Community structure of the *Cimex* microbial associates**

While the core microbiome of *C. lectularius* individuals collected from urban Cincinnati is dominated by two OTUs from Proteobacteria (*Wolbachia* and the unnamed γ-proteobacterium), their relative abundances differed among populations (Fig 2A). *Wolbachia* was found to be highly abundant (84-85%) in populations 5 and 8, respectively, whereas the unnamed γ-proteobacterium was found at only 11-13% in these populations. In contrast, populations 6 and 3 exhibited slightly more of the unnamed γ-proteobacterium (56 and 49%) than *Wolbachia* (40 and 43%) (Fig 2A). Populations 2, 4, 7 and 9 showed similar relative abundances of these two genera on average (~60% *Wolbachia* and ~34% unnamed γ-proteobacterium).

We found a strong and highly significant negative correlation between the abundance of OTU 1 (*Wolbachia*) versus that of OTU 2 (unnamed γ-proteobacterium)—individuals that exhibited a higher abundance of OTU 1 tended to show a lower abundance of OTU 2 and *vice versa* ($r = -0.87$, $P = <0.0001$; Fig 3A). This pattern was also apparent when OTU abundance was averaged according to population ($r = -0.94$, $P = 0.0005$; Fig 3B). Since populations varied significantly according to weight ($F_7 = 14.96$, $P < 0.0001$; SFig 1), we hypothesized that the variation in the relative abundances of OTU 1 and OTU 2 might be related to the size of the insects within populations. Such an effect could be expected if the individuals within populations were of similar age, and, the different populations were at different stages of development when collected. Variation in OTU abundance related to insect weight might also be expected if a particular bacterial species provided a growth benefit relative to another bacterial species.

We uncovered a potential trend for a moderate, positive relationship between the log-transformed abundance of OTU 1 and log-transformed insect weight (g) ($r = 0.34$, $P = $ 0.092), and, a significant negative correlation between the log-transformed abundance of OTU 2 and insect weight (r = -0.44, P = 0.028). We then performed an analysis of covariance separately for each of the two dominant OTUs, using individual weight as a covariate in the model to determine if the inclusion of weight explained more variation in OTU abundance than the effect of population. We uncovered no evidence that variation in weight explained the log-transformed abundance of OTU 2 (Table 3A), as only the population effect was significant in this model. OTU 1, on the other hand, exhibited a significant weight by population interaction (Table 3B) suggesting that OTU 1 abundance varied in relation to weight differently among the different populations. Because a significant interaction between the covariate and an independent variable in an ANCOVA violates the assumption of homogeneity of slopes between treatments (in our case, population), the other factors in the model are considered unreliable. However, this model, like that of OTU 2, provides no evidence that weight as a covariate explains more variation in abundance of the OTU than does the effect of population. Thus, the population of origin appears to be a stronger indicator of relative OTU abundance than does insect weight for OTU 2 and possibly OTU 1.

The next 8 most abundant OTUs—most of which were detected in almost all populations but at very low frequency (Table 1)—did not follow similar patterns as the highly abundant OTUs. An OTU classified as *Shinella*, from the Rhizobiales order, was found in population 3 at 1.63% and yet at only 0.2% in population 6 and at lower frequency in all other populations (Fig 2B). An OTU from the Frankineae order was present at relatively similar frequency among populations 5, 6, and 7 (~0.68%). These same populations also exhibited a relatively similar frequency of an OTU that was another member of the Enterobacteriaceae. The majority of our bootstrap values from the RDP classifier provided 100% support, especially for the high-frequency OTUs, with the taxonomic assignment of a few low-frequency OTUs between 80-90% (see STable2).

To determine if the patterns of abundance represent significant microbial community structuring across populations, we estimated the Yue and Clayton similarity coefficient (θ)[37] among the individual *Cimex* microbiomes. The Yue and Clayton θ index provides a similarity measure based on species proportions, whether shared or not between communities, and ranges from 0 to 1, with 1 = complete similarity and 0 = complete dissimilarity. We then visualized the resulting distance matrices using a principle components analysis (PCoA) and performed an analysis of molecular variance (AMOVA) to determine if the spatial separation in the PCoA plot was statistically significant, i.e., individuals of the same population shared similar communities and at similar abundances. The PCoA plot (Fig 4; SMovie1) of this metric of community similarity shows evidence of population structure in that individual microbiomes from the same population tended to cluster with one another; further, the AMOVA uncovered evidence that the variation among population spatial structure was greater than variation between individuals collected from the same population ($F_7 = 11.25$, $P = 0.008$).

**Patterns of microbial diversity among the *Cimex* microbiomes**

On average, 22.56 ± 2.48 (SE) OTUs were uncovered across the populations, with four libraries either approaching or exhibiting approximately 60 OTUs (Fig 5). On the lower end, three libraries appeared to asymptote between 10-15 OTUs. Rarefaction analysis shows that individual *Cimex* microbiomes vary in richness (Fig 5), and that the libraries were sampled at different depths. Those that were not sampled to completion could harbor rare and potentially important bacterial taxa. Coverage of the *Cimex* microbiomes averaged at 98.5%, suggesting that, on average, 1.5 unique OTUs would be expected every additional 100 sequences.

Indices of diversity and evenness varied significantly among populations (Fig 6; STables 3A&B, STable 4), and this variation appeared to be driven largely by the relative low diversity and evenness of populations 5 and 8. These two populations were dominated by *Wolbachia* relative to the unnamed γ-proteobacterium, and exhibited an average of 18.5 (population 5) and 14.3 (population 8) OTUs. Rarefaction analysis of the Inverse Simpson index shows that unequal sampling among *Cimex* libraries did not excessively influence this estimate of diversity (SFig 2).

## Discussion

### The core *Cimex* microbiome

We found two dominant OTUs to make up 97% of the *C. lectularius* microbiome. The RDP classifier assigned these OTUs to the α-proteobacteria *Wolbachia*, an obligate endosymbiont, and the unnamed γ-proteobacterium, a putative primary or secondary endosymbiont from the γ-proteobacteria. Both taxa have been described from previous surveys of *C. lectularius*—they are transovarially transmitted, persist in bacteriocytes (both taxa [29,33,38] and/or are found in many cell types of the bed bug (unnamed γ-proteobacterium [33]). *Wolbachia* comprised the largest fraction of our microbiome survey (~63%) and was present in all populations and individuals. Broader geographical assessments have previously indicated that *Wolbachia* infections are present in ~95% of surveyed *C. lectularius* populations [32], and earlier experimental work suggested a role for this symbiont in *C. lectularius* fertility [39,40,41]. More recently, it has been shown that the presence of *Wolbachia* in bed bugs provides a fertility benefit through the production of B vitamins [33] supporting the idea that this taxon is an obligate nutritional mutualist of the bed bug. Thus, the high frequency of *Wolbachia* uncovered in this study is relatively unsurprising, but provides evidence that our short-read 16S rDNA assessment is largely congruent with expectations.

While different sub-types of *Wolbachia* are known to inhabit many arthropods and can range in association with their hosts along a parasitic to mutualistic continuum [42,43,44], there is comparatively less known about the life-style of the second most common OTU in our dataset, the unnamed γ-proteobacterium. A blastn analysis of a representative sequence from this OTU shows high similarity (99%, e-value = $3e^{-139}$) to a previously identified maternally transmitted γ-proteobacteria endosymbiont from *C. lectularius* called BEV-like endosymbiont [29,33]. Phylogenetic assessment of this taxon found it to exhibit greater than 99% similarity to BEV [29], a parasitic bacteria reported from the leafhopper *Euscelidius variegatus* [45] and closely related to the soft-rot plant pathogenic genera *Dickeya*, *Pectobacterium*, and *Erwinia* [46]. Close phylogenetic relationships between endosymbionts found in plant-feeding and blood-feeding insect hosts have been uncovered in other facultative endosymbiont groups—*Rickettsia*, *Sodalis* and *Arsenophonus* [47,48,49]—and represent putative examples of horizontal transfer, potentially due to their ability to be cultivated and persist outside of hosts [46]. While the phenotypic effects of this γ-proteobacteria on *C. lectularius* are largely unknown, there are accounts of pen-strep treatment (which does not affect *Wolbachia*) leading to reductions in

264  egg production in *C. lectularius*, suggesting that the removal of the γ-proteobacteria can
265  lead to reductions in fertility [38] and thus it may prove to be a nutritional mutualist as
266  well.
267
268  **Trade-off in taxon abundance in the *Cimex* core microbiome**
269       A very striking result from this comprehensive survey of the *C. lectularius*
270  microbiome is the strong negative correlation between the abundance of *Wolbachia* and
271  the unnamed γ-proteobacterium. Such a relationship could be due to functional, historical,
272  developmental or sex-biased reasons. For example, we wondered if, due to *Wolbachia*'s
273  role as a nutritional mutualist [33], populations and/or individuals with higher abundance
274  of *Wolbachia* might exhibit higher fitness then populations with lower abundance of this
275  bacterial group. While we did not assess the reproductive fitness of these wild-collected
276  populations, we did measure each individual's weight prior to DNA extraction. Including
277  weight as a covariate in an ANCOVA did not uncover evidence that the size of the insect was
278  a predictor of OTU abundance, thus, to the extent that insect weight is a correlate to an
279  individual's fitness, we do not have evidence that a higher abundance of one OTU could
280  provide a greater functional benefit than the other. In the same vein, we do not have
281  evidence that a higher abundance of OTU 2, the unnamed γ-proteobacterium, provides a
282  fitness detriment to the insects, as measured by their weight. This perhaps supports the
283  suggestion that the unnamed γ-proteobacterium provides *C. lectularius* a fitness benefit
284  similarly to *Wolbachia* (and as suggested by [38]), or at the very least be considered
285  commensalistic. However, the relative importance of each endosymbiont should be assessed in a more comprehensive and empirical manner before such conclusions can be drawn.

Other potential explanations for the negative correlation in abundance of these major taxa could be that populations exhibit sex-ratio bias or were collected at different stages in development. The average sex ratio across all populations was 0.684±0.057 (Avg female:male ratio ± SE), which did not covary with the abundance of either OTU among populations (data not shown). Thus, the negative correlation in abundance between OTUs 1 and 2 is not likely to be driven by differences in sex among the populations. We attempted to use adult individuals from each population for DNA extraction such that targeted individuals were ~5 mm in size. However, since we used field-collected individuals for our survey, which we did not culture in the lab, we cannot guarantee that the size of the insects indicates life stage. Significant microbiome community structure has been identified across life history stages of mosquito and subcortical beetles [50,51], thus, an assessment of microbiome community structure changes as related to *Cimex* metamorphosis, and in relation to timing of blood meals, would be informative.

Finally, the negative correlation between the two taxa could be due to historical patterns of infection. This hypothesis is best supported by the current data—although weight was not found to be a predictor for OTU abundance of the two dominant types, we did uncover a highly significant effect of population of origin. Recent work shows that the genetic structure of *C. lectularius* is high across surveyed populations from the eastern US, with little variation within populations [52]. This pattern suggests multiple introductions of *C. lectularius* into the United States—and hence high diversity among populations— followed by subsequent inbreeding within populations due to the establishment of the population by a single mated female [52]. Our results of significant population structure in the microbiome of *C. lectularius* supports this pattern since the dominant members of the microbiome are maternally transmitted, and thus the establishment of a new population *via* a single female would establish a maternally-inherited microbiome. That we uncovered community structure in the microbiomes among populations is even more striking in light of our experimental decision to assess the whole-insect microbiome, *i.e.,* environmental and endosymbiont bacteria together.

As presented above, the negative correlation in abundance of these two dominant OTUs could simply be due to historical patterns of co-infection and establishment of populations such that manipulation of the relative abundance of these two OTUs within a host would have no effect on host fitness. Alternatively, these microbial abundance patterns could reflect historical coevolution between populations of *C. lectularius* and their microbiomes. This is practically relevant in light of the suggestion to manipulate *Wolbachia* as a mechanism for biocontrol of bed bugs [33]. If the presence of *Wolbachia* acts to prevent the establishment of other, putatively human-pathogenic bacterial species in *C. lectularius* (a scenario of colonization resistance [2]), targeting *Wolbachia* for the biocontrol of bed bugs could potentially lead to unwanted effects, such as the effective vectoring of pathogenic microbes or viruses (reviewed in [3]). Although there is no solid evidence to date that bed bugs can vector human disease, that we see a trade-off between the abundance of endosymbionts in our survey of geographically close populations suggests that the potential for interaction between these endosymbiont species within *C. lectularius* deserves further attention.

**Identification of putative human pathogens in *Cimex***

One of the goals of this survey was to determine if human pathogenic bacteria could be detected among populations of *C. lectularius* collected in a dense urban setting. Because *C. lectularius* colonies infest walls and other cracks and crevices of structures, live in high density in their excrement, and shuttle around dust, the environmental microbes (i.e. those that can be found on the exoskeleton) were of interest as were potential pathogenic endosymbionts. Although bed bugs have not been definitively shown to transmit disease, 45 human pathogens with the potential for transmission through bed bug vectors, most likely mechanical in contrast to true vector, have been previously identified within or on the insects and/or their excrement [24]. Furthermore, bed bug bites and exposure to bed bug excrement can trigger systemic reactions such as asthma and anaphylaxis [25,26]. Our survey of the whole insect microbiome uncovered at least 5 genera to which known or putative human pathogen species belong. For example, two OTUs of *Bacillus* were detected in three of our eight populations surveyed—other blood-feeding arthropods (mosquitos, horse and deer flies) have been documented to vector *B. anthracis*, the bacteria that causes anthrax [53]. We also found two OTUs in one population from the family that *Coxiella burnetti* (Q-fever) belongs to, along with two *Streptococcus* OTUs found in five populations and one *Staphylococcus* OTU found in four populations. While the use of short-read 16S rDNA sequences does not enable identification of the bacterial species and strains of interest to human health, the use of such technology across a broad area, followed by either 16S rDNA Sanger sequencing and/or high-depth metagenomic sequencing could identify human pathogens at the level of the species.

**Similar bacterial lineages across populations suggest host-microbe association**

Our distance-based screen of 54,411 sequences from 31 individual *Cimex* libraries uncovered 359 OTUs—a number that is high relative to other surveys of arthropod bacterial communities (e.g., 139 OTUs found across species and wild populations of *Drosophila*, [9]; 74 OTUs uncovered across 11 Drosophila populations, [17]; both Sanger sequenced, long-read surveys). Our use of multiple individuals collected across geographically separate populations, the method of sample preparation, and 454 sequencing error are all factors that likely influenced this high number. To reduce sequencing error on our estimates, we followed the standard operating procedure from the Schloss lab (http://www.mothur.org/wiki/Main_Page), which included sequence quality checks and trimming of low-quality sequence and homopolymers, as well as a pre-clustering step that bins rare sequences that are within 2 bp of a more common OTU into the more common group. We also searched for and removed chimeric sequences within the database—such screens are reported to reduce the overall chimera rate to 1% [54]. Further, we utilized techniques that reduce PCR amplification errors such as using a high-fidelity Taq polymerase and the pooling of triplicate amplifications of each individual's DNA. With these checks in place, we still uncovered a large number of OTUs that were very low-copy (239 OTUs represented by 1 or 2 sequences). We removed these low-copy sequences from the overall dataset and performed each step of the presented analysis with OTUs represented by three or more sequences. This highly conservative treatment of the data uncovered 49 OTUs at the 3% level of divergence—many fewer than the 359 uncovered with the full dataset. However, all of our major conclusions were the same: populations exhibited variable diversity and significant microbial community structuring, we found a similar relative abundance of the 10 most abundant OTUs and a highly significant negative correlation between the two most abundant OTUs (data not shown). Although we cannot distinguish rare, environmental OTUs from 454 sequencing error with our full dataset, if we were to assume an error rate of 1% from non-identified chimeras, and another 18% error from 454 sequencing chemistry (as identified using the V4 region in [55]), we would reduce our number of OTUs by 68 to 291. Thus, it is reasonable to predict that many of the low-copy OTUs from across eight populations are from environmental sources, such as bacteria picked up and transported on the surface of the individuals.

In comparison, the OTUs that we uncovered across all or most individuals and/or populations, and are represented by more than a few sequences, are likely to be either endosymbionts or consistently associated with the *C. lectularius* exoskeleton rather than strictly environmental. Ten such OTUs were uncovered in all populations, and 11 OTUs were found in 15 or more individuals, albeit most at very low frequency. Of note, *Methylobacterium* was found in 25 of the individual libraries and in all populations—this group of bacteria, often associated with plants, is vectored by glassy-winged sharpshooters [56] and thus has been suggested as a sharpshooter biocontrol agent. An OTU classified as from the *Sphingomonas* genus, a known endosymbiont of ticks [10], was uncovered in 22 individual bed bug libraries. Finally, in 21 libraries, we detected the presence of *Propionibacterium*, a genus of bacteria to which *P. acnes* belongs—this bacteria, which causes acne in humans, is often found on skin and infecting human blood [57,58].

**Variation among populations in the structure, diversity and evenness of the *Cimex* microbiome**

The *Cimex* microbial populations exhibited significant community structuring as well as variation in diversity and evenness. Whereas most of the Inverse Simpson values range from 1.8-2.1 among populations, populations 5 and 8 show comparatively lower values, from ~1.3 to 1.4. These results are likely influenced by the relative abundances of the most abundant OTUs—*Wolbachia* dominated these two populations, whereas the other populations exhibited less extreme relative abundances of *Wolbachia* versus the unnamed γ-proteobacterium. Further, populations 5 and 8 did not exhibit wide differences in the abundance of the low-copy OTUs that were similarly uncovered in most populations, suggesting that the significant variation among populations in diversity and evenness are driven largely by the differences in abundance of *Wolbachia* and the unnamed γ-proteobacterium. A survey of the geographical distribution of the microbial populations from natural *Drosophila* populations has also uncovered population variation in community composition and taxon richness [17], and further, there is evidence that the major taxa of the tick microbiome show heterogeneity among populations [10]. This survey of the *Cimex* microbiome is relatively unique compared to previous assessments in that between 2 and 5 individuals were assessed separately per population. The use of multiple sequenced microbiomes per population thus allows us to definitely test for the presence of population structure among wild-caught *C. lectularius* populations, which has implications for understanding endosymbiont patterns of co-occurrence and for generating expectations regarding the frequency of human pathogenic bacteria.

**Conclusions**

The data presented here provide a novel view of variation in the abundance of bed bug endosymbiont bacteria across populations, and an indication that the two dominant endosymbiont bacteria may interact within hosts or host lineages. These patterns suggest a range of interesting hypotheses, such as: Do the two dominant members of the bed bug microbiome competitively exclude one another, or, does host fitness depend on the presence of both types in an additive fashion? Are the patterns of population structure and relative abundance of the two dominant groups from naturally collected populations strictly due to inheritance from a founding maternal individual, or might there be unexamined environmental influences that are responsible for the negative correlation in the abundance of the two types? How might these patterns of community structure impact potential attempts to eradicate or control bed bug infestations? Learning more about the interaction of the two dominant members across multiple established bed bug populations will have broad implications for our understanding of bed bug physiology, human health, and bed bug biocontrol.

## Methods

### Sample collection and DNA isolation

*C. lectularius* populations were collected from residential locations in the Cincinnati area by Scherzinger Pest Control and generously donated to the Baucom lab for this project (see Fig 1 for locations). Technicians recorded the perceived severity of each infestation, which ranged from 5 (most severe) to 1 (least severe). Most donated populations were scored '5,' with the exception of a single population that was scored as a '3' (see STable 1 for specific information). The population sizes were not otherwise characterized. Information identifying the exact location of each residence was not divulged by the pest company in accordance with their individual contracts with residents; however, a general address was supplied for our records. Other identifying information about each residence, such as whether it was from an apartment building or single-family residence was not provided. At least ten individuals per population were randomly collected from the residence and placed in sterile 50 mL falcon tubes. Once collected, each population was starved for ~7 days before being snap frozen in liquid nitrogen and stored at -20 °C until DNA isolation. Five individuals from each of the eight populations were randomly selected and DNA was isolated from each individual using the Powersoil DNA Isolation Kit (MO Bio, Carlsbad, CA, USA) and tested for purity on a NanoDrop 2000 Spectrophotometer (ThermoScientific, Waltham, MA, USA). Across populations, average DNA yield was 3.97 (±0.62 SE) ng/uL with an average 260/280 score of 2.30 (±0.11 SE). Individuals were not surface sterilized prior to DNA isolation since the whole insect microbial community, and the community the insects potentially transport on their exteriors and/or in their feces was of interest.

**16S Library and Sequencing**

16S rDNA genes were amplified using the bacterial universal primers 530F (5'-GTGCCAGCMGCNGCGG-3') and 1100R (5'-GGGTTNCGNTCGTTG-3') which amplified ~ 600 bp corresponding to the V4-V6 hypervariable region of *E. coli*. DNA from each individual was amplified by PCR in order to create amplicon libraries for pyrosequencing using Titanium GS FLX chemistry on a 454 sequencing machine. Primers thus included the Titanium Fusion A or B primers and a 15 bp error-correcting mid-tag optimized for individual amplicon library identification from a single pyrosequencing run (Roche, Penzberg, DE); the general sequence of each primer was LinkerA-midtag-530F and LinkerB-1100R (see STable2 for specific primers utilized in this study). The HotStar HiFidelity Polymerase Kit (Qiagen, Valencia, CA, USA) was used for PCR under the following conditions: 95 °C for 5 min followed by 40 cycles of 94 °C for 15 s; 60 °C for 1 min; 72 °C for 1 min; and a final elongation of 72 °C for 10 min. Each sample was amplified in triplicate, pooled and cleaned using the QIAquick PCR Purification Kit (Qiagen, Valencia, CA, USA) followed by normalization of each library to 25 ng using the SequalPrep Normalization Plate Kit (Invitrogen, Carlsbad, CA, USA). Once normalized, the 48 libraries were pooled and sent to the Purdue Genomics Lab for pyrosequencing according to standard protocols (Roche, Penzberg, Germany).

**Sequence analysis**

Sequence quality control was performed using mothur [36] in the following manner: first, the trim.seqs command was used to trim the sequence when the average quality score over a 50 bp window dropped below 35, and sequences with ambiguous bases (N),

484  homopolymers (>8 bases), or sequences that were <200 bp were removed. The pdiffs=2
485  and bdiffs=1 options of this command were utilized in order to allow a maximum of 2 and 1
486  nucleotide differences in the primer and barcode sequences, respectively. Sequences were
487  aligned to the SILVA-compatible alignment database using align.seqs, and then trimmed to
488  a common region. To further reduce the influence of pyrosequencing error on downstream
489  analysis, sequences were clustered using the pre.cluster command, and those that were
490  within 2 bp of similarity to a more abundant sequence were merged with the more
491  abundant type. Next, potential chimeric sequences were removed from the full dataset by
492  utilizing the SILVA gold database as the reference in the chimera.uchime command. Finally,
493  the classify.seqs and remove.lineage commands were utilized to identify and remove
494  potential mitochondrial and chloroplast contaminants.
495
496  Bacterial taxonomy was assigned to each sequence in the improved dataset using the
497  classify.seqs command, which implements the Naïve Bayesian classifier of RDP, as
498  implemented in mother [36]. Following taxonomic assignment, sequences were assigned to
499  OTUs using the cluster command, following which the consensus taxonomy for each OTU at
500  the 3% level of divergence was determined using the classify.otu command [36]. The
501  abundance of each $OTU_{0.03}$ within each individual and population was determined using the
502  shared.otu command. Broad patterns of the presence or absence of specific OTUs across all
503  *Cimex* populations were considered utilizing the improved, full dataset. To compare
504  estimates of diversity and community structure among populations, we first normalized
505  the data such that 1000 sequences were randomly subsampled from each individual.

Individuals with <1000 sequences following quality control were thus excluded from further analyses (Table S1).

Commands within the program mothur were used to produce rarefaction curves and assess both alpha- and beta-diversity of the communities. The coverage, number of observed OTUs$_{0.03}$, the Shannon evenness and the inverse Simpson statistics were calculated for each individual, and an ANOVA using Type I SS was performed in the R statistical programming language (http://www.R-project.org) to determine if populations varied in these estimates of microbial community diversity. To describe the community dissimilarity, (and conversely, similarity in community membership and structure across populations) the Yue and Clayton theta statistic was estimated via the dist.shared command. The resulting distance matrix was visualized in a principle components ordination plot using the first two axes, and then a rotation of the 3D matrices was performed using the ggobi package in R [59].

We next performed an analysis of molecular variance using the AMOVA command of mothur to determine if there was significant structure among populations, followed by the metastats command to identify OTUs$_{0.03}$ that were differentially represented in the separate *Cimex* populations. The presence of a core microbiome was assessed by first using the make.shared command to determine which OTUs were shared across all individuals and by then visualizing the abundance of OTUs using the heatmap.2 command of R (http://www.R-project.org).

**Acknowledgements**: We thank E. Buschbeck for assistance identifying the sex of the bed bug individuals and providing general advice on culturing an arthropod, P. Boland from Scherzinger Pest Control for performing the collections and providing the samples, and V. Corby-Harris and members of the Baucom lab for critiques that helped improve a previous version of this manuscript. Funding was provided by the University of Cincinnati.

683
684
685
686
687
688
689
690
691
692
693
694
695
696
697
698
699
700

**Figure Legends**

Fig. 1 Map of *Cimex lectularius* populations collected from the greater Cincinnati metropolitan area in Cincinnati, OH, USA.

Fig. 2 Composition and abundance of the *C. lectularius* microbiome within eight wild populations collected from Cincinnati,OH, USA. Each individual *C. lectularius* library was randomly subsampled to 1000 sequences, and the abundance of each OTU was averaged according to population. Note that population 6 is represented by a single individual. A. Relative abundance of the two most abundant bacterial families and genera within the populations. B. Relative abundance of the next 8 most abundant bacterial families and genera. Note the difference in scale in the respective keys for panels A and B.

Fig. 3 The relationship between the abundance of OTU1 (*Wolbachia*) and OTU2 (unclassified), in **A.** each individual *C. lectularius* library, and **B.** averaged according to *C. lectularius* population.

Fig. 4 Principle component analysis of the *Cimex* microbiome. The Yue and Clayton measure of dissimilarity between the structures of the microbiome communities (θ-YC) was estimated and visualized using the dist.shared and pcoa commands of mothur (Schloss et al 2009).

Fig. 5 Rarefaction analysis of observed richness of the microbiome of *C. lectularius* individuals. All calculations were performed with mothur (Schloss et al 2009). OTUs were defined at the 3% level of sequence divergence. Individuals collected from the same population are represented by the same color.

Fig. 6 Boxplots of the diversity and evenness of the bacterial communities found in association with *C. lectularius* individuals, averaged according to population. Individual libraries were randomly subsampled to 1000 sequences each prior to estimating diversity parameters. Calculations were performed with mothur (Schloss et al 2009) at the 3% level of sequence divergence. The number of individuals within each population is presented in STable 1.

**Supplementary Figure Legends**

SFig. 1 A boxplot showing population variation in insect weight (g). Black lines represent the population median, the top and bottom of the boxes show the 75 and 25% quartile, respectively, and the whiskers show the maximum and minimum values of weight.

SFig. 2 Rarefaction of the Inverse Simpson index for each *Cimex* library.

SMovie 1 A rotating 3D display of the Yue and Clayton measure of dissimilarity between the structures of the microbiome communities (θ-YC), implemented by ggobi in R.

Table 1. The % abundance of the 10 most abundant OTUs shared across all or most populations. The number of individual *C. lectularius* (or libraries) that exhibit each OTU is presented along with the number of populations with the OTU.

| OTU ID Number | Phylum/Class/Family | Genus | Abundance | Number of libraries OTU present | Number of populations OTU present |
|---|---|---|---|---|---|
| | **Actinobacteria** | | | | |
| | Actinobacteria | | | | |
| 29 | Frankineae | Unclassified | 0.23 | 18 | 7 |
| 4 | Propionibacterineae | Unclassified | 0.11 | 21 | 8 |
| 31 | Streptomycineae | Unclassified | 0.07 | 17 | 7 |
| | **Proteobacteria** | | | | |
| | Alphaproteobacteria | | | | |
| 1 | Anaplasmataceae | *Wolbachia* | 63.48 | 31 | 8 |
| 26 | Shinella | *Unclassified* | 0.22 | 16 | 8 |
| 23 | Sphingomonadaceae | *Sphingomonas* | 0.13 | 22 | 8 |
| 15 | Methylobacteriaceae | *Methylobacterium* | 0.10 | 25 | 8 |
| | Gammaproteobacteria | | | | |
| 2 | Enterobacteriaceae | Unclassified | 33.42 | 31 | 8 |
| 85 | Pseudomonadaceae | *Pseudomonas* | 0.16 | 16 | 8 |
| 11 | Enterobacteriaceae | Unclassified | 0.15 | 20 | 8 |

| Genus detected in this study | Number of OTUs identified | Pathogen previously detected in *C. lectularius* or *C. hemipterus*[1] |
| --- | --- | --- |
| *Bacillus* | 2 | *Bacillus anthracis* |
| *Coxiellaceae* | 2 | *Coxiella burnetii* (Q fever) |
| *Staphylococcus* | 1 | *Staphylococcus aureus* |
| *Streptococcus* | 2 | *Streptococcus pneumonia* |
| *Wolbachia* | 3 | *Wolbachia* spp |

Table 2. Genera detected in the wild populations of *C. lectularius* previously detected in either wild or laboratory strains of *C. lectularius* ([1]from Delaunay et al 2011).

| A | Source of variation | df | SS | F | P-value |
|---|---|---|---|---|---|
| | Population | 7 | 1.144 | 12.59 | 0.0003 |
| | Weight | 1 | 0.012 | 0.95 | 0.3525 |
| | Population × Weight | 6 | 0.122 | 1.57 | 0.2418 |
| | Residual | 10 | 0.129 | | |

| B | Source of variation | df | SS | F | P-value |
|---|---|---|---|---|---|
| | Population | 7 | 0.437 | 8.81 | 0.0014 |
| | Weight | 1 | 0.008 | 1.24 | 0.2921 |
| | Population × Weight | 6 | 0.155 | 3.64 | 0.0352 |
| | Residual | 10 | 0.071 | | |

Table 3. Results of an ANCOVA assessing the effect of population and insect weight on variation in the abundance of **A.** OTU 2 (Unclassified) and **B.** OTU 1 (*Wolbachia*). Insect weight was used as the covariate in each analysis.

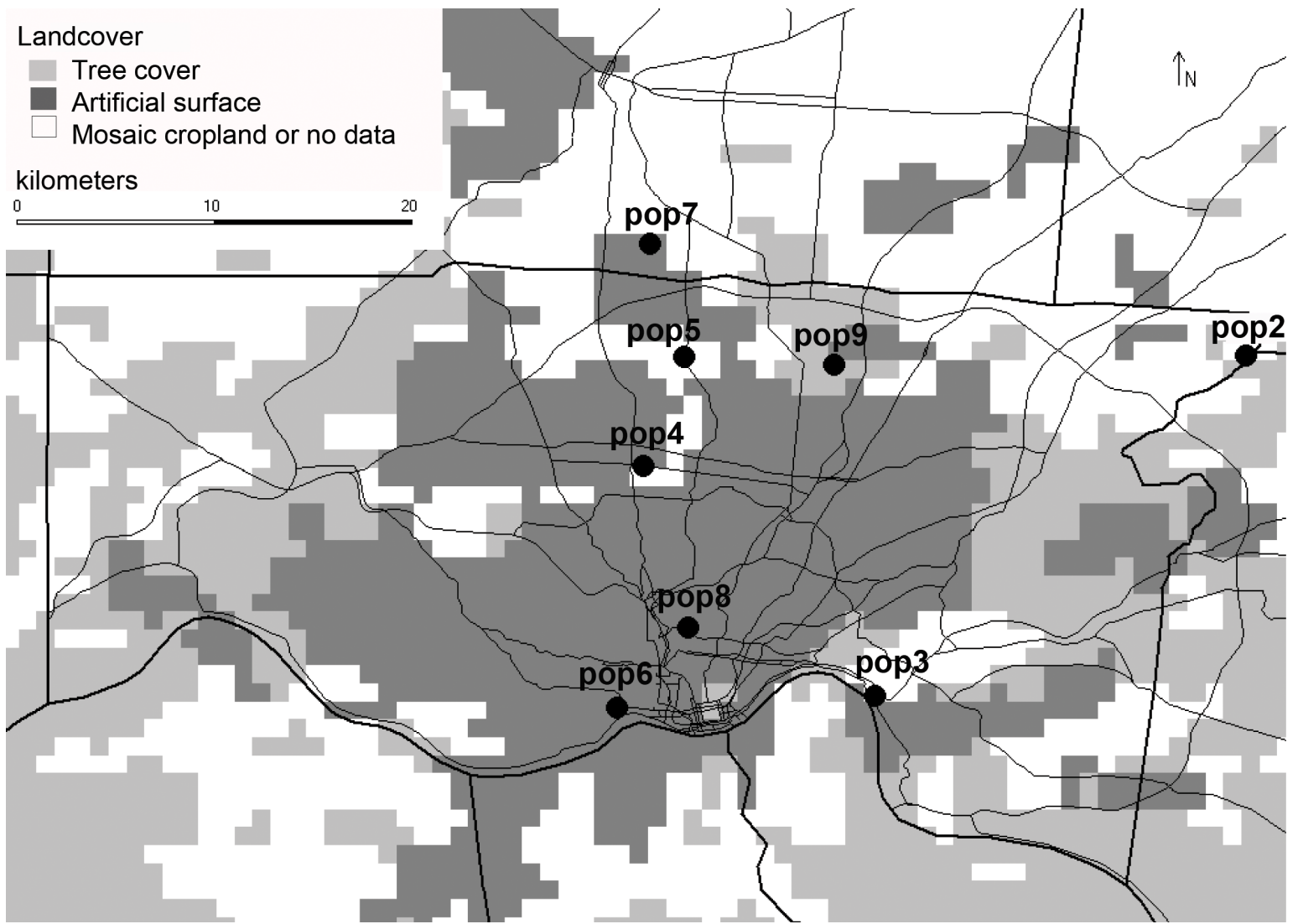

Fig 1.

A

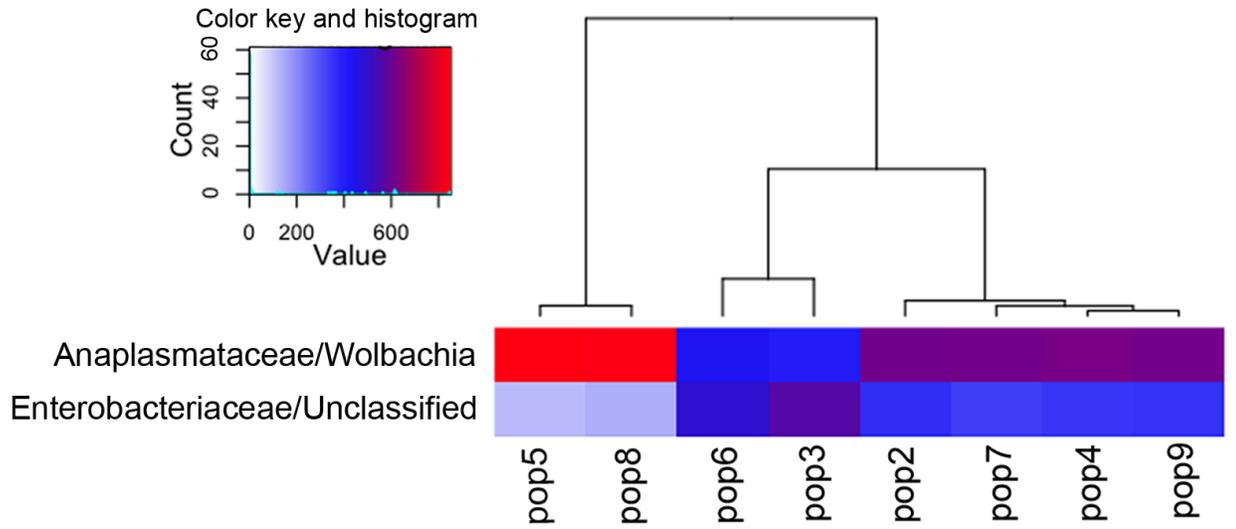

B

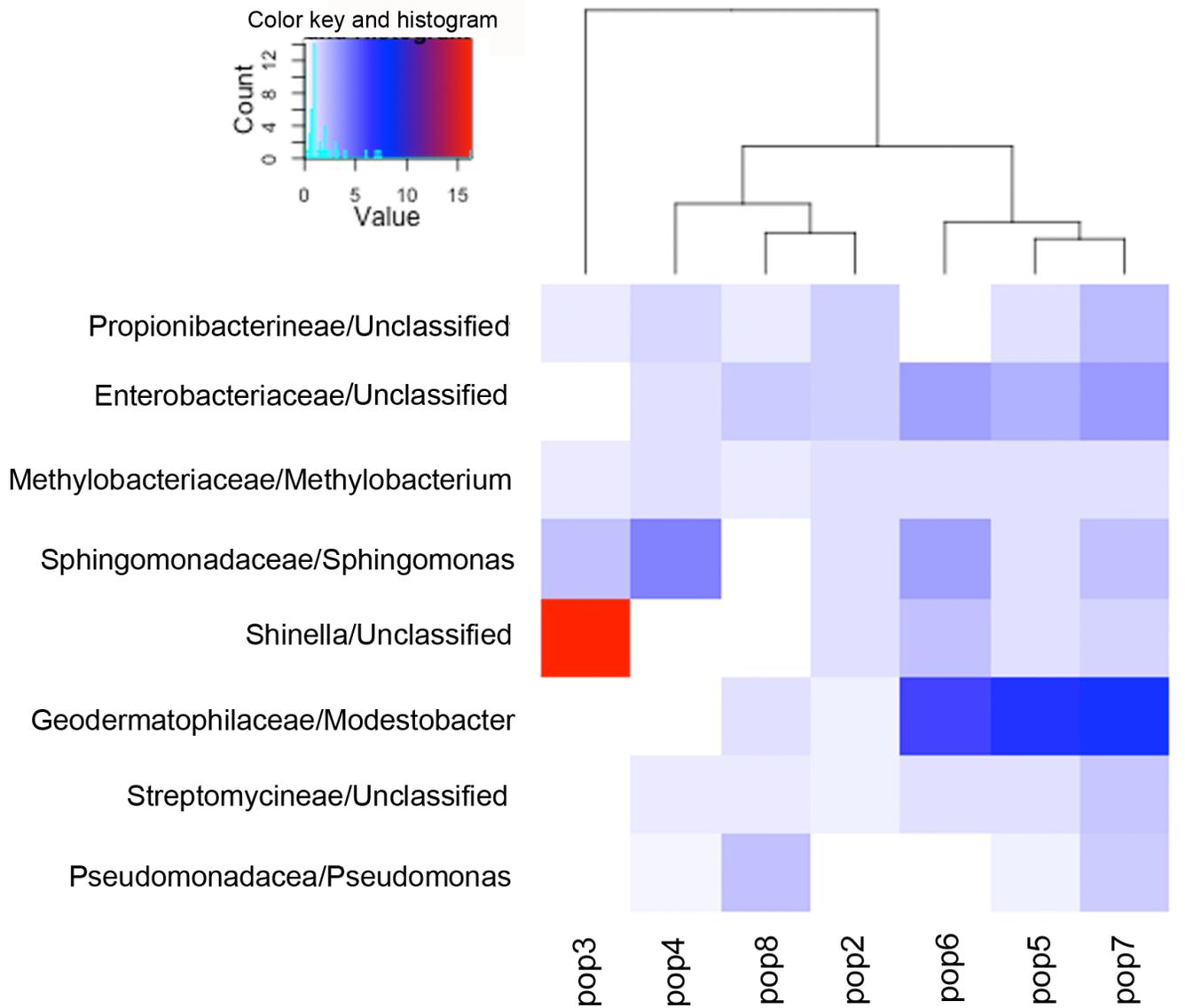

Fig 2.

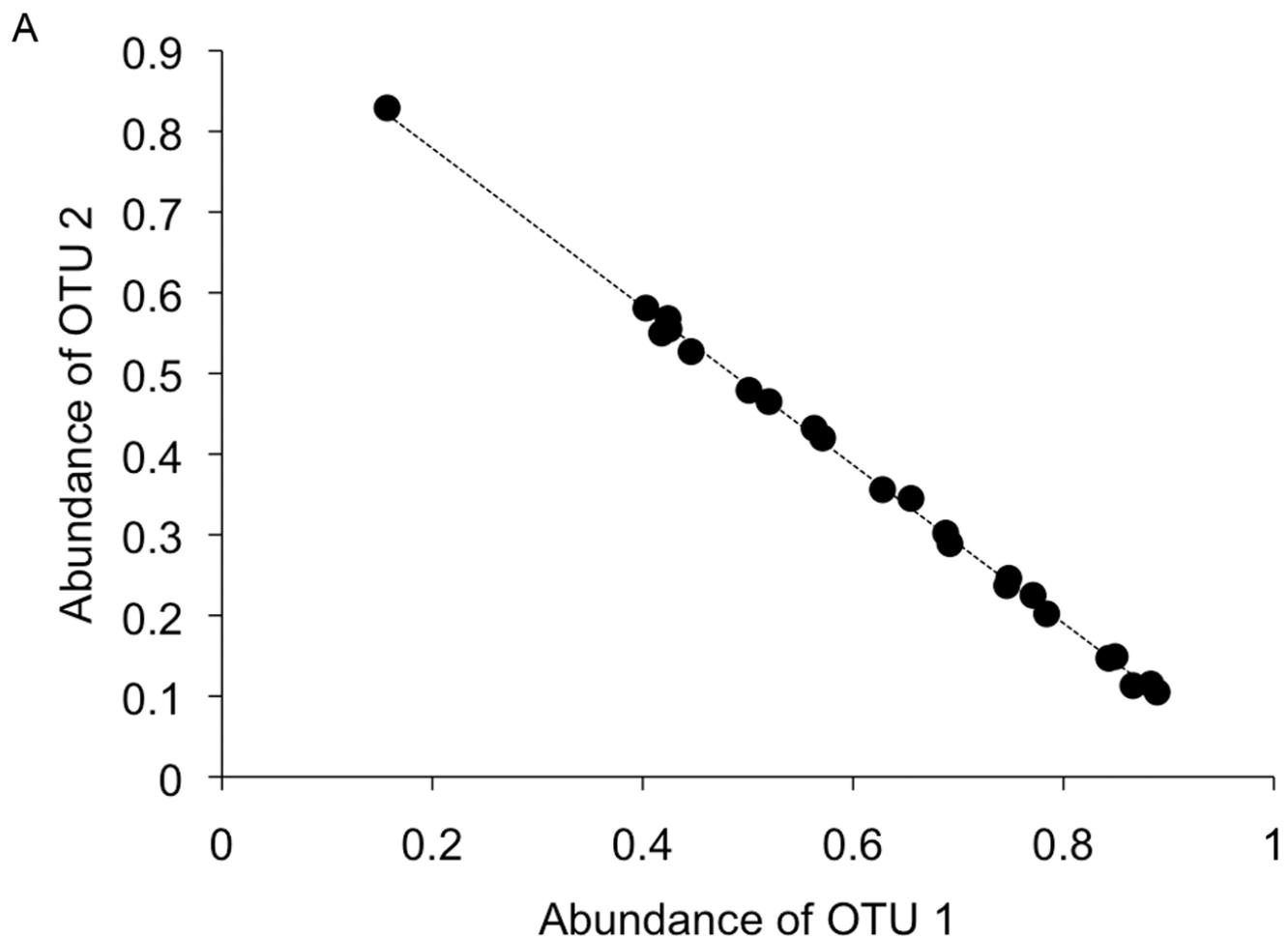

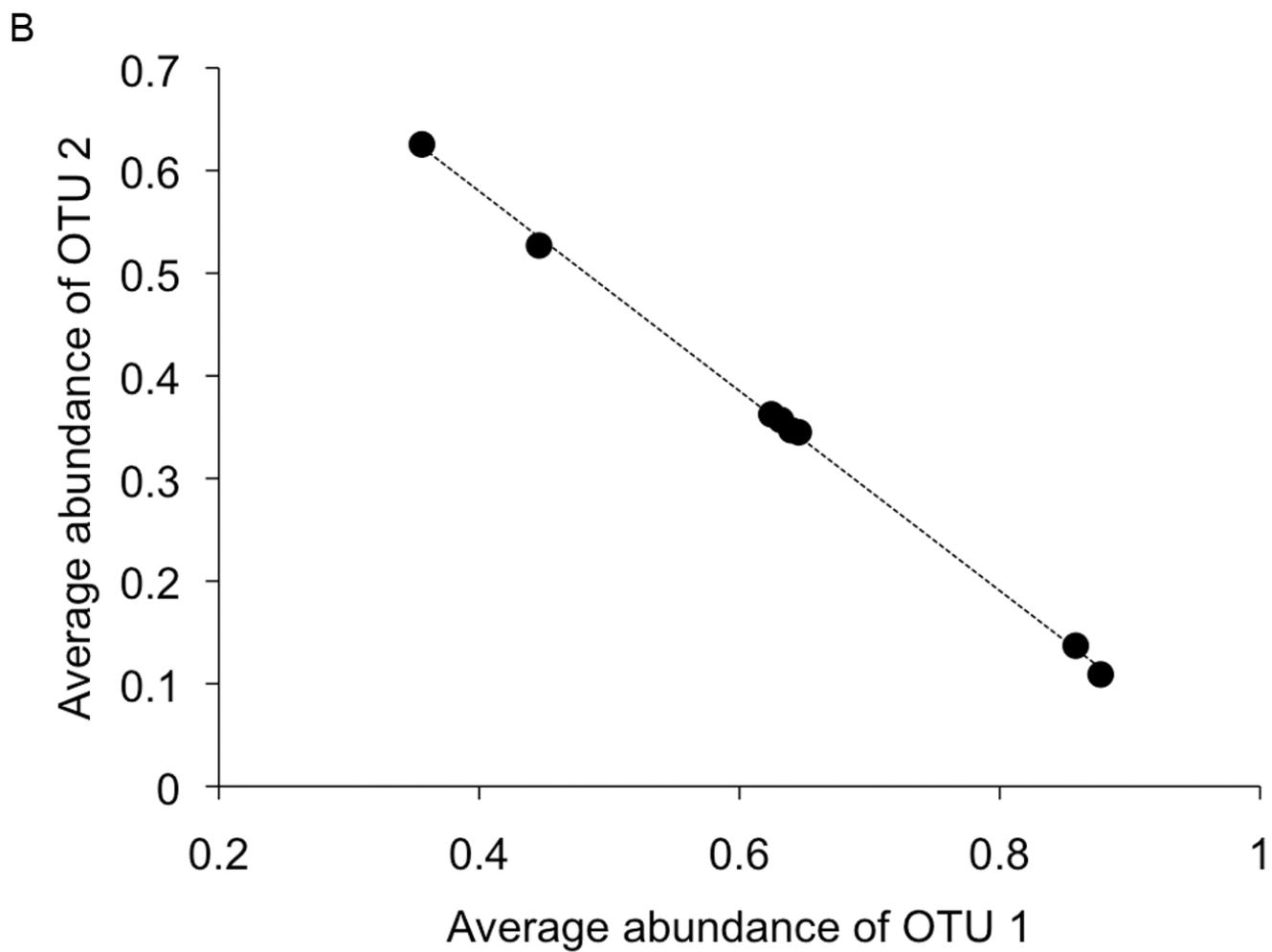

Fig 3

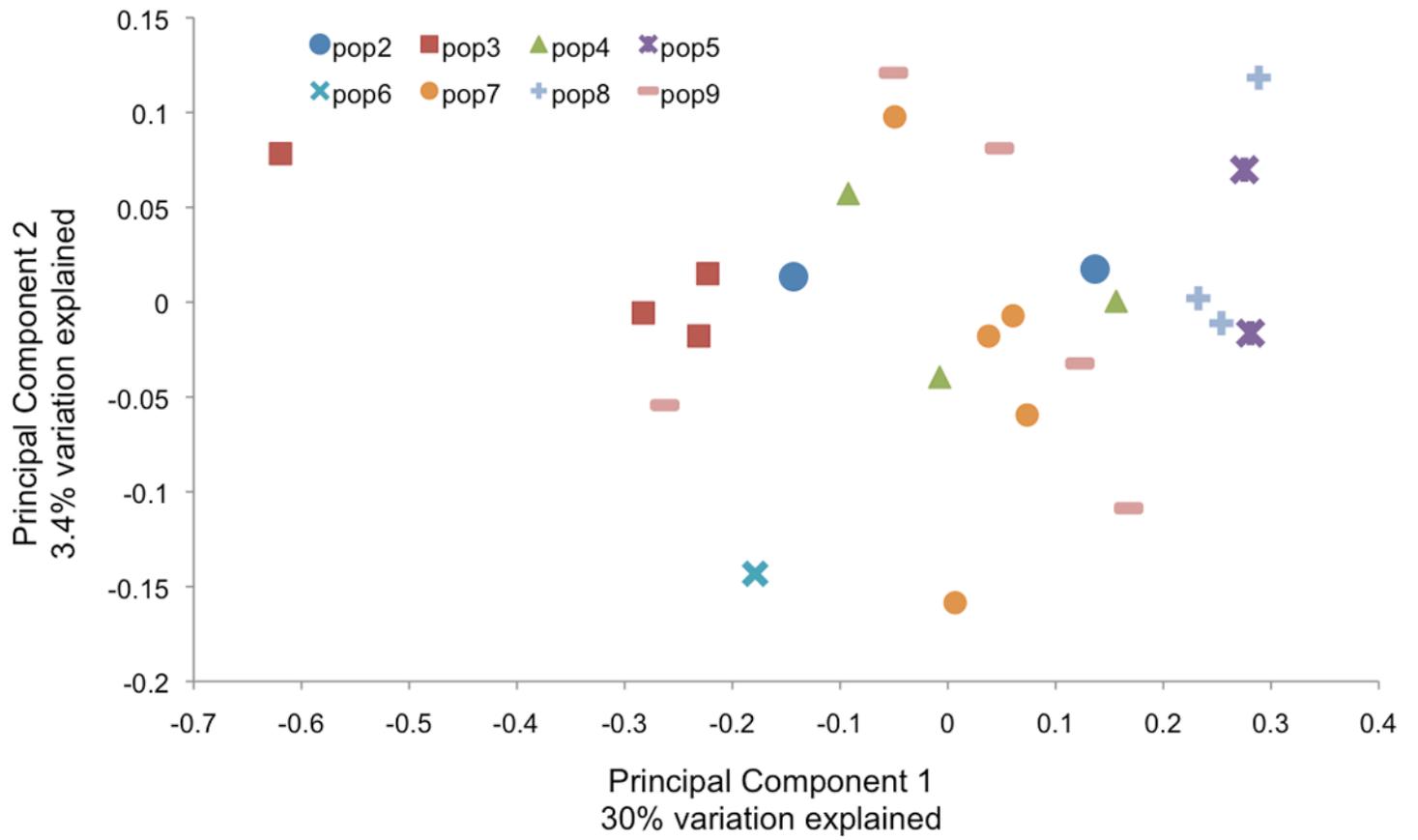

Fig 4

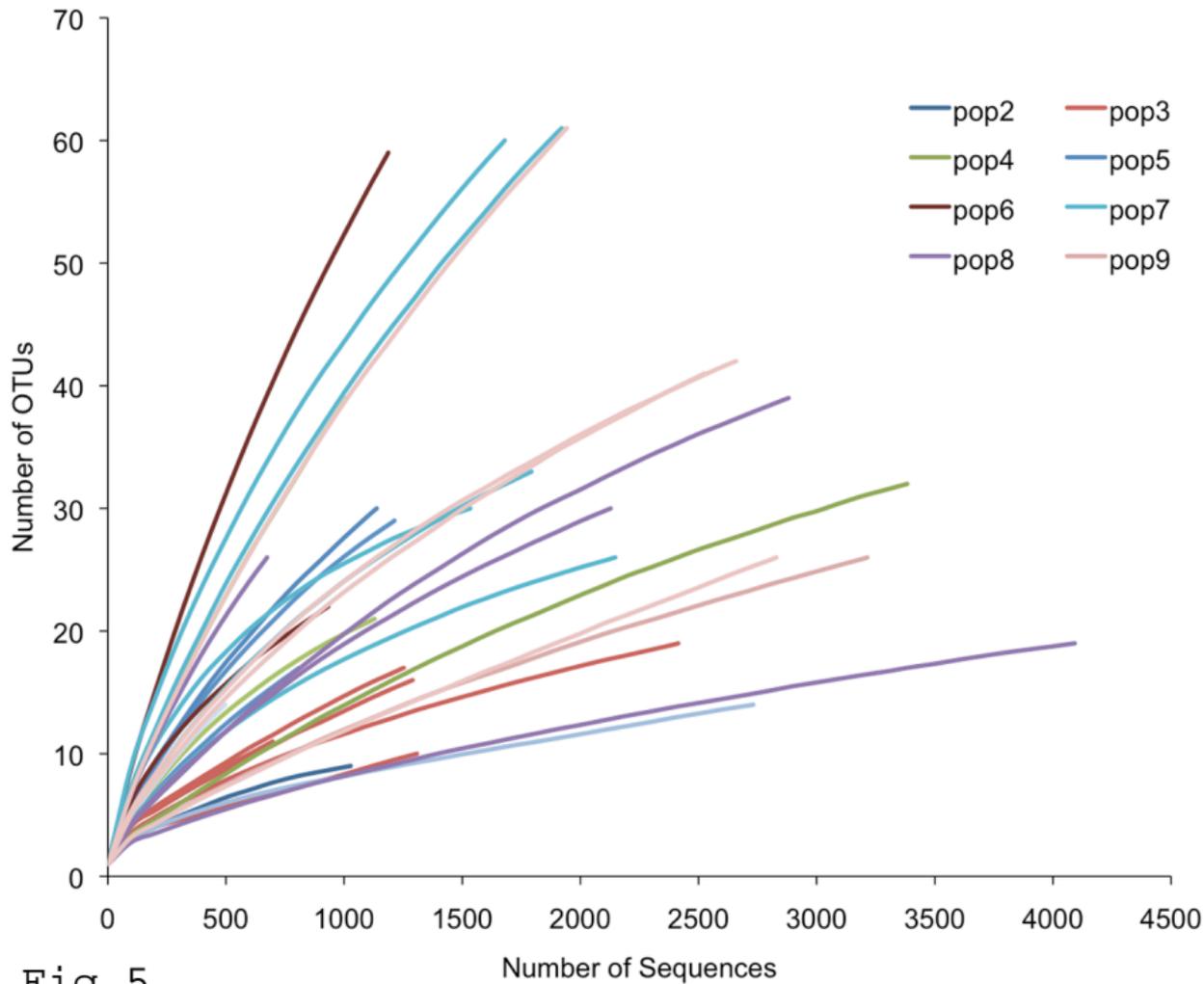

Fig 5

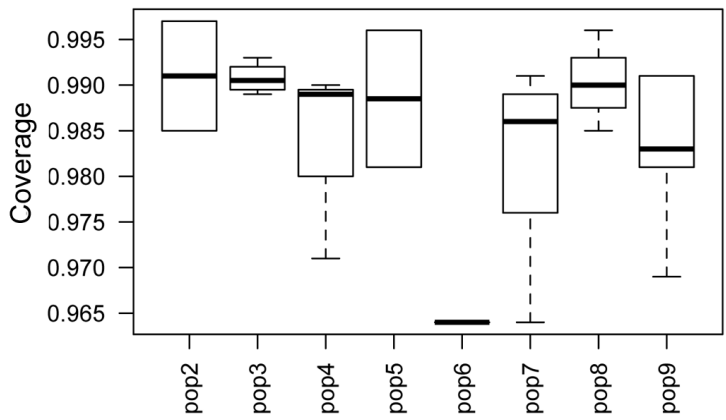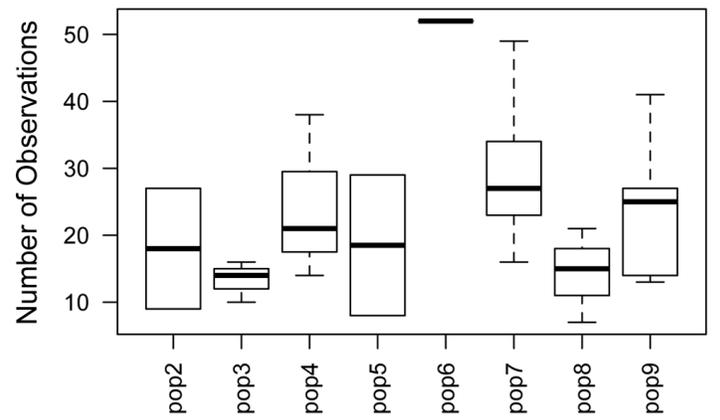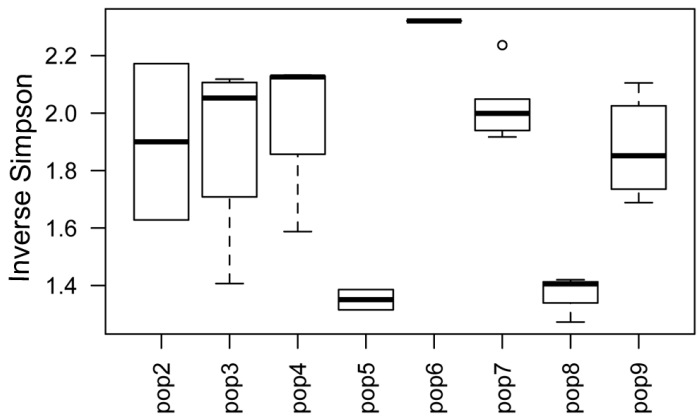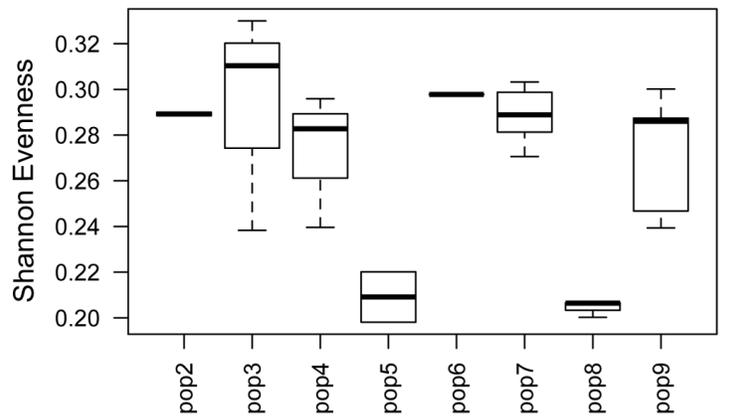

Fig 6

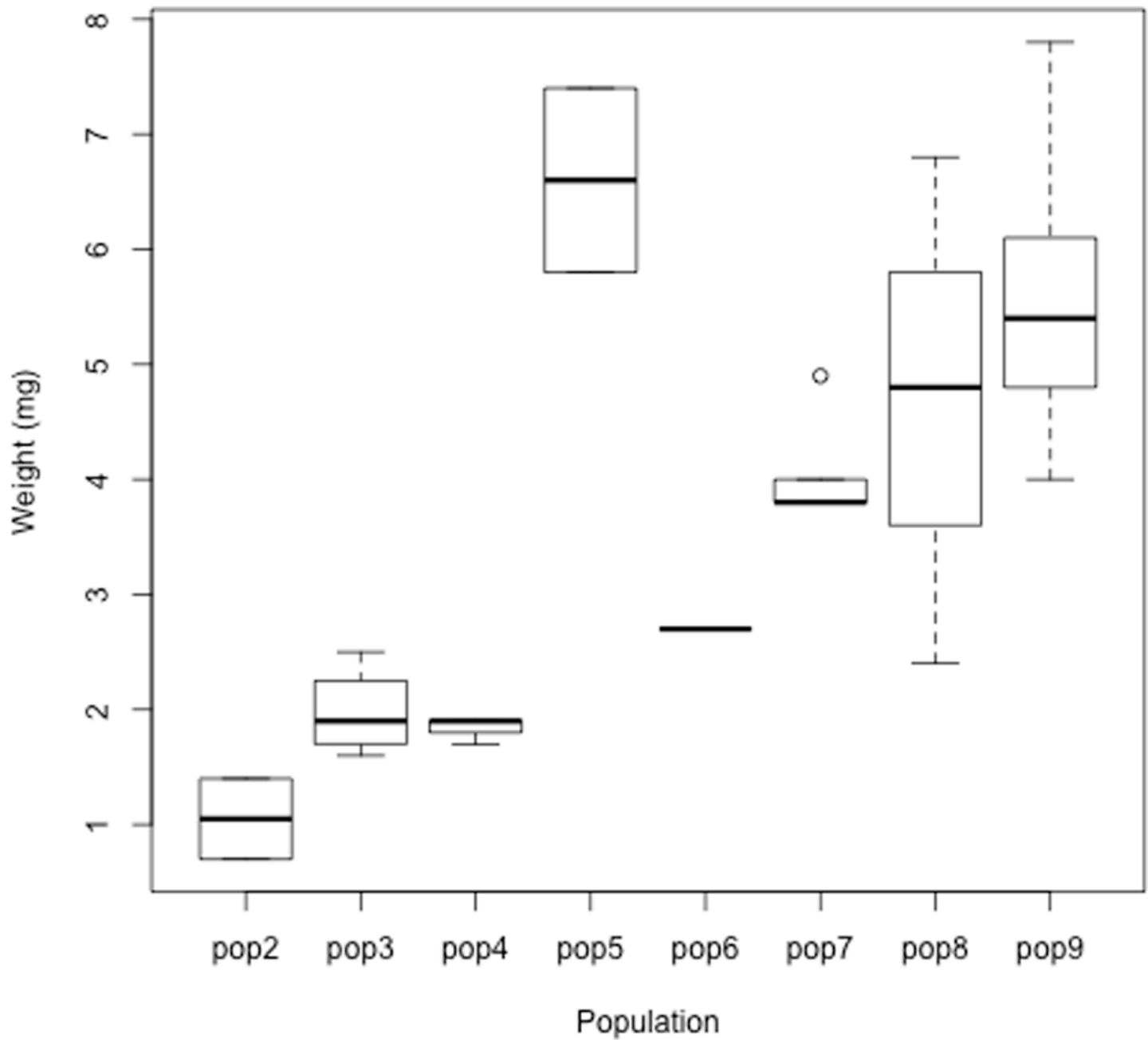

SFig1

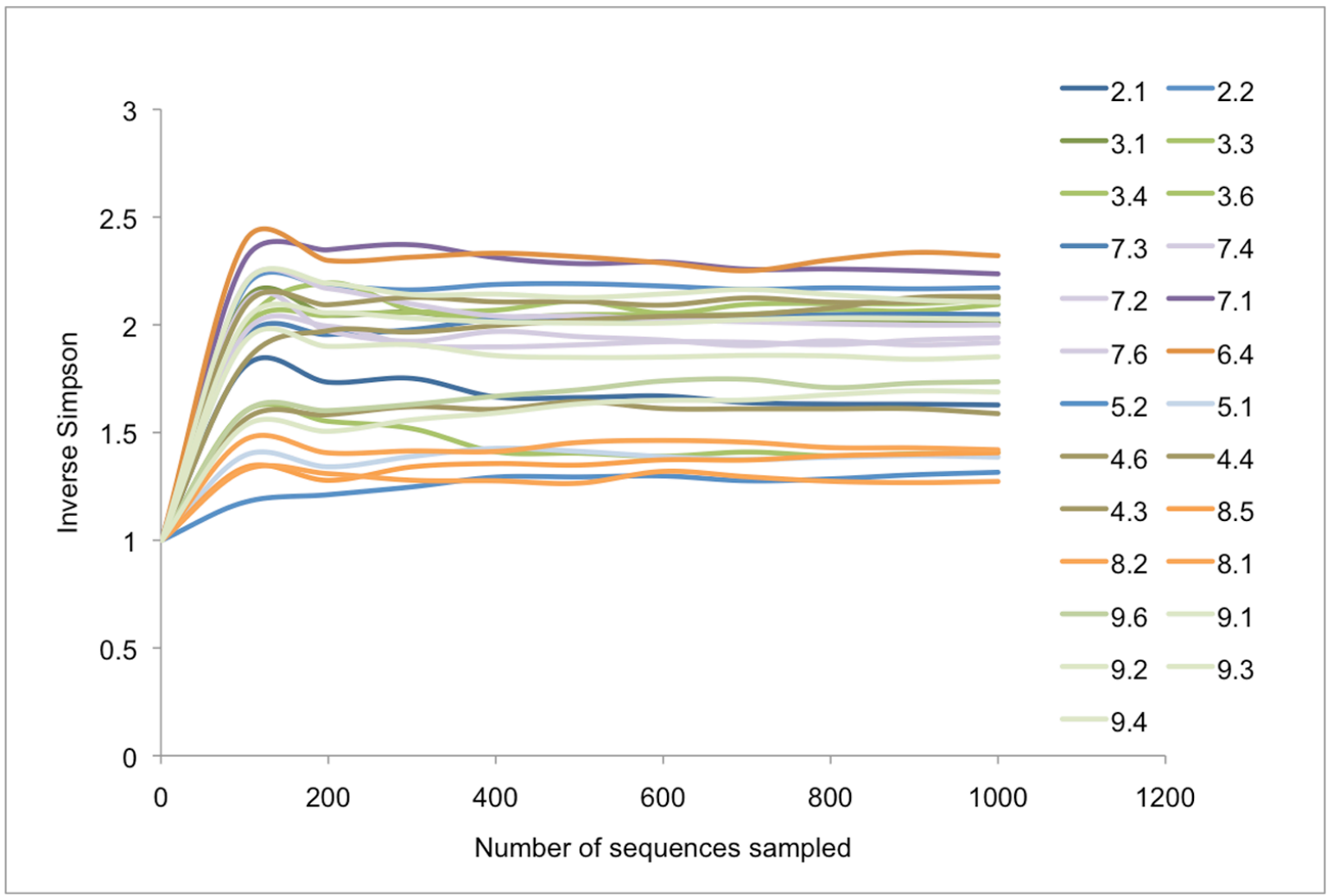

SFig 2